# Octonic Electrodynamics

**Victor L. Mironov**[*] **and Sergey V. Mironov**

Institute for physics of microstructures RAS,
603950, Nizhniy Novgorod, Russia

In this paper we present eight-component values "octons", generating associative noncommutative algebra. It is shown that the electromagnetic field in a vacuum can be described by a generalized octonic equation, which leads both to the wave equations for potentials and fields and to the system of Maxwell's equations. The octonic algebra allows one to perform compact combined calculations simultaneously with scalars, vectors, pseudoscalars and pseudovectors. Examples of such calculations are demonstrated by deriving the relations for energy, momentum and Lorentz invariants of the electromagnetic field. The generalized octonic equation for electromagnetic field in a matter is formulated.

## I. INTRODUCTION

Hypercomplex numbers[1-4] especially quaternions are widely used in relativistic mechanics, electrodynamics, quantum mechanics and quantum field theory[3-10] (see also the bibliographical review Ref. 11). The structure of quaternions with four components (scalar and vector) corresponds to the relativistic space-time structure, which allows one to realize the quaternionic generalization of quantum mechanics[5-8]. However quaternions do not include pseudoscalar and pseudovector components. Therefore for describing all types of physical values the eight-component hypercomplex numbers enclosing scalars, vectors, pseudoscalars and pseudovectors are more appropriate.

The idea of applying the eight-component hypercomplex numbers for the description of the electromagnetic field is quite natural since Maxwell's equations are the system of four equations for scalar, vector, pseudoscalar and pseudovector values. There are a lot of papers, that describe the attempts to realize representations of Maxwell equations using different eight-component hypercomplex numbers such as biquaternions[4,12,13], octonions[14-16] and multivectors generating the associative Clifford algebras[17-19]. However all considered systems of hypercomplex numbers do not have a consistent vector interpretation, which leads to difficulties in the description of vectorial electromagnetic fields.

This paper is devoted to describing electromagnetic fields on the basis of eight-component values "octons", which generate associative noncommutative algebra and have the clearly defined, simple geometric sense. The paper has the following structure. In section 2 we consider the peculiarities of the eight-component octonic algebra. In section 3 the generalized octonic equations for the electromagnetic field in a vacuum are formulated. In section 4 the derivations of the relations for energy, momentum and Lorentz invariants of electromagnetic field are demonstrated. At last in section 5 we consider the generalized octonic equations for the electromagnetic field in a matter.

---

[*] E-mail: mironov@ipmras.ru

## II. ALGEBRA OF OCTONS

The values of four types (scalars, vectors, pseudoscalars and pseudovectors) differing with respect to spatial inversion are used for the description of the electromagnetic field. All these values can be integrated into one spatial object. For this purpose in the present paper we propose the special eight-component values, which will be named "octons" for short.

The eight-component octon $\breve{G}$ is defined by the following expression

$$\breve{G} = c_0 e_0 + c_1 e_1 + c_2 e_2 + c_3 e_3 + d_0 a_0 + d_1 a_1 + d_2 a_2 + d_3 a_3, \qquad (1)$$

where $e_0 \equiv 1$, values $e_1$, $e_2$, $e_3$ are axial unit vectors (pseudovectors), $a_0$ is the pseudoscalar unit and $a_1$, $a_2$, $a_3$ are polar unit vectors. The octonic components $c_n$ and $d_n$ ($n = 0, 1, 2, 3$) are numbers (complex in general). Thus the octon is the sum of a scalar, vector, pseudoscalar and pseudovector. The full octon basis is

$$e_0, \ e_1, \ e_2, \ e_3, \ a_0, \ a_1, \ a_2, \ a_3. \qquad (2)$$

The rules for multiplication of polar and axial basis vectors are formulated taking into account the symmetry of their products with respect to the operation of spatial inversion. For polar unit vectors $a_k$ ($k = 1, 2, 3$) the following rules of multiplication take place:

$$a_k^2 = 1, \qquad (3)$$

$$a_j a_k = -a_k a_j \ (\text{for } j \neq k, \ j = 1, 2, 3). \qquad (4)$$

The conditions (4) describe the property of noncommutativity for vector product. The same rules are defined for axial unit vectors multiplication:

$$e_k^2 = 1, \qquad (5)$$

$$e_j e_k = -e_k e_j \ (\text{for } j \neq k). \qquad (6)$$

The rules (3) - (6) allow one to represent the length square of any polar or axial vector as the sum of squares of its components. We emphasize that the square of vector length is positively defined. The rules (3) and (5) lead to some special requirements for the vector product in octonic algebra. Let $a_1$, $a_2$, $a_3$ and $e_1$, $e_2$, $e_3$ be the right Cartesian bases and corresponding unit vectors are parallel each other. Taking into consideration (3)-(5) and the fact that the product of two different polar vectors is an axial vector, we can represent the rules for cross multiplication of polar unit vectors in the following way:

$$a_1 a_2 = i e_3, \ a_2 a_3 = i e_1, \ a_3 a_1 = i e_2, \qquad (7)$$

where $i$ is the imaginary unit. Then the rules of multiplication for axial basis units can be written

$$e_1 e_2 = i e_3, \ e_2 e_3 = i e_1, \ e_3 e_1 = i e_2. \qquad (8)$$

Let us define the pseudoscalar unit $a_0$ as the product of parallel unit vectors corresponding to the different bases:

$$a_0 = a_k e_k. \qquad (9)$$

Squaring (9) we can see that $a_0^2 = 1$. Note that the unit $a_0$ commutates with each unit vector.

Summarized commutation and multiplication rules are represented in the table 1.

*Table 1. The rules of multiplication and commutation for the octon's unit vectors.*

|       | $e_1$   | $e_2$   | $e_3$   | $a_0$  | $a_1$   | $a_2$   | $a_3$   |
|-------|---------|---------|---------|--------|---------|---------|---------|
| $e_1$ | *1*     | $ie_3$  | $-ie_2$ | $a_1$  | $a_0$   | $ia_3$  | $-ia_2$ |
| $e_2$ | $-ie_3$ | *1*     | $ie_1$  | $a_2$  | $-ia_3$ | $a_0$   | $ia_1$  |
| $e_3$ | $ie_2$  | $-ie_1$ | *1*     | $a_3$  | $ia_2$  | $-ia_1$ | $a_0$   |
| $a_0$ | $a_1$   | $a_2$   | $a_3$   | *1*    | $e_1$   | $e_2$   | $e_3$   |
| $a_1$ | $a_0$   | $ia_3$  | $-ia_2$ | $e_1$  | *1*     | $ie_3$  | $-ie_2$ |
| $a_2$ | $-ia_3$ | $a_0$   | $ia_1$  | $e_2$  | $-ie_3$ | *1*     | $ie_1$  |
| $a_3$ | $ia_2$  | $-ia_1$ | $a_0$   | $e_3$  | $ie_2$  | $-ie_1$ | *1*     |

We would like to emphasize especially that octonic algebra is associative. The property of associativity follows directly from multiplication rules.

Thus the octon $\breve{G}$ (1) is the sum of the scalar value $c_0$, the pseudovector value (axial vector) $\vec{c} = c_1 \boldsymbol{e_1} + c_2 \boldsymbol{e_2} + c_3 \boldsymbol{e_3}$, the pseudoscalar value $\tilde{d}_0 = d_0 \boldsymbol{a_0}$ and the vector value (polar vector) $\vec{d} = d_1 \boldsymbol{a_1} + d_2 \boldsymbol{a_2} + d_3 \boldsymbol{a_3}$:

$$\breve{G} = c_0 + \vec{c} + \tilde{d}_0 + \vec{d} \ .$$

Hereinafter octons will be indicated by the "∪" symbol, pseudovectors by a double arrow "↔", pseudoscalars by a wave "∼" and vectors by an arrow "→". The values $c_k$ and $d_k$ ($k = 1, 2, 3$) are the projections of the axial vector $\vec{c}$ and the polar vector $\vec{d}$ on the corresponding unit vectors directions. Note, that equality of two octons means the equality of all corresponding components.

Let us consider the rules of multiplication of two octons in detail. First, the result of octonic multiplication of two polar vectors $\vec{c}_1$ and $\vec{c}_2$ is the sum of scalar and pseudovector values:

$$\vec{c}_1 \vec{c}_2 = \{c_{11}\boldsymbol{a_1} + c_{12}\boldsymbol{a_2} + c_{13}\boldsymbol{a_3}\}\{c_{21}\boldsymbol{a_1} + c_{22}\boldsymbol{a_2} + c_{23}\boldsymbol{a_3}\} = $$
$$= \{c_{11}c_{21} + c_{12}c_{22} + c_{13}c_{23}\} + i\{c_{12}c_{23} - c_{13}c_{22}\}\boldsymbol{e_1} + i\{c_{13}c_{21} - c_{11}c_{23}\}\boldsymbol{e_2} + i\{c_{11}c_{22} - c_{12}c_{21}\}\boldsymbol{e_3} \ . \quad (10)$$

Hereinafter we will denote the scalar multiplication (internal product) by the symbol "·" and round brackets:

$$(\vec{c}_1 \cdot \vec{c}_2) = c_{11}c_{21} + c_{12}c_{22} + c_{13}c_{23} \ ,$$
$$(\vec{d}_1 \cdot \vec{d}_2) = d_{11}d_{21} + d_{12}d_{22} + d_{13}d_{23} \ ,$$
$$(\vec{c} \cdot \vec{d}) = \{c_1 d_1 + c_2 d_2 + c_3 d_3\}\boldsymbol{a_0} \ .$$

Vector multiplication (external product) will be denoted by the symbol "×" and square brackets:

$$[\vec{c}_1 \times \vec{c}_2] = i\{c_{12}c_{23} - c_{13}c_{22}\}\boldsymbol{e_1} + i\{c_{13}c_{21} - c_{11}c_{23}\}\boldsymbol{e_2} + i\{c_{11}c_{22} - c_{12}c_{21}\}\boldsymbol{e_3} \ ,$$
$$[\vec{d}_1 \times \vec{d}_2] = i\{d_{12}d_{23} - d_{13}d_{22}\}\boldsymbol{e_1} + i\{d_{13}d_{21} - d_{11}d_{23}\}\boldsymbol{e_2} + i\{d_{11}d_{22} - d_{12}d_{21}\}\boldsymbol{e_3} \ ,$$

$$[\vec{c} \times \vec{d}] = i\{c_2 d_3 - c_3 d_2\}\boldsymbol{a_1} + i\{c_3 d_1 - c_1 d_3\}\boldsymbol{a_2} + i\{c_1 d_2 - c_2 d_1\}\boldsymbol{a_3}.$$

In all other cases round and square brackets will be used for the priority definition. Thus taking into account the considered designations, the octonic product of two vectors (10) can be represented as the sum of scalar and vector products

$$\vec{c}_1 \vec{c}_2 = (\vec{c}_1 \cdot \vec{c}_2) + [\vec{c}_1 \times \vec{c}_2].$$

Then the product of two octons can be represented in the following form

$$\breve{G}_1 \breve{G}_2 = \{c_{10} + \vec{c}_1 + \tilde{d}_{10} + \vec{d}_1\}\{c_{20} + \vec{c}_2 + \tilde{d}_{20} + \vec{d}_2\} =$$
$$= c_{10}c_{20} + c_{10}\vec{c}_2 + c_{10}\tilde{d}_{20} + c_{10}\vec{d}_2 + c_{20}\vec{c}_1 + (\vec{c}_1 \cdot \vec{c}_2) + [\vec{c}_1 \times \vec{c}_2] + \tilde{d}_{20}\vec{c}_1 + (\vec{c}_1 \cdot \vec{d}_2) + [\vec{c}_1 \times \vec{d}_2] +$$
$$+ \tilde{d}_{10}c_{20} + \tilde{d}_{10}\vec{c}_2 + \tilde{d}_{10}\tilde{d}_{20} + \tilde{d}_{10}\vec{d}_2 + c_{20}\vec{d}_1 + (\vec{d}_1 \cdot \vec{c}_2) + [\vec{d}_1 \times \vec{c}_2] + \tilde{d}_{20}\vec{d}_1 + (\vec{d}_1 \cdot \vec{d}_2) + [\vec{d}_1 \times \vec{d}_2].$$

We can indicate some connection between octonic algebra and algebra of quaternions. Indeed it is easy to see that the rules of octonic multiplication and commutation take place for the values based on the quaternionic imaginary units $q_k$ ($k = 1, 2, 3$; $q_k^2 = -1$)

$$\boldsymbol{e_k} = i q_k, \qquad \boldsymbol{a_k} = i q_k \boldsymbol{a_0},$$

but it needs the introduction of a new (nonquaternionic) pseudoscalar element $\boldsymbol{a_0}$.

In conclusion in this section we would like to note that formally the algebra of octons can be considered as the variant of complexified Clifford algebra. However in contrast to Clifford algebra the octonic unit vectors $\boldsymbol{a_1}$, $\boldsymbol{a_2}$, $\boldsymbol{a_3}$ and $\boldsymbol{e_1}$, $\boldsymbol{e_2}$, $\boldsymbol{e_3}$ are the real true vectors but not complex numbers. In this connection octons have a clear well-defined space-geometry sense.

### III. OCTONIC FORM OF ELECTRODYNAMICS EQUATIONS

The octonic algebra can be naturally applied to the description of the electromagnetic field in a vacuum. The potential of the electromagnetic field is represented as an incomplete four-component octon

$$\breve{\Pi} = \varphi + \vec{A} = \varphi + A_1 \boldsymbol{a_1} + A_2 \boldsymbol{a_2} + A_3 \boldsymbol{a_3},$$

where $\varphi$ is the scalar potential, $\vec{A}$ is the vector potential. The four-component current also can be defined as an incomplete octon

$$\breve{J} = 4\pi\rho + \frac{4\pi}{c}\vec{j} = 4\pi\rho + \frac{4\pi}{c}(j_1 \boldsymbol{a_1} + j_2 \boldsymbol{a_2} + j_3 \boldsymbol{a_3}).$$

Then using the octonic differentiation operator

$$\hat{P} = \left(\frac{1}{c}\frac{\partial}{\partial t} + \vec{\nabla}\right) = \left(\frac{1}{c}\frac{\partial}{\partial t} + \frac{\partial}{\partial x_1}\boldsymbol{a_1} + \frac{\partial}{\partial x_2}\boldsymbol{a_2} + \frac{\partial}{\partial x_3}\boldsymbol{a_3}\right)$$

and conjugated operator

$$\hat{P}^+ = \left(\frac{1}{c}\frac{\partial}{\partial t} - \vec{\nabla}\right) = \left(\frac{1}{c}\frac{\partial}{\partial t} - \frac{\partial}{\partial x_1}\boldsymbol{a_1} - \frac{\partial}{\partial x_2}\boldsymbol{a_2} - \frac{\partial}{\partial x_3}\boldsymbol{a_3}\right),$$

we can write the generalized equation of electrodynamics in the compact octonic form

$$\hat{P}^+ \hat{P}\, \breve{\Pi} = \breve{J}. \tag{11}$$

Indeed multiplying $\hat{P}^+$ and $\hat{P}$ operators in (11), we obtain the wave equation for potentials of electromagnetic field in the form

$$\left(\frac{1}{c^2}\frac{\partial^2}{\partial t^2} - \Delta - \left[\vec{\nabla}\times\vec{\nabla}\right]\right)\breve{\Pi} = \breve{J} \ . \tag{12}$$

For the potentials described by twice differentiable functions $\left[\vec{\nabla}\times\vec{\nabla}\right]\breve{\Pi} = 0$, and equation (12) becomes

$$\left(\frac{1}{c^2}\frac{\partial^2}{\partial t^2} - \Delta\right)\breve{\Pi} = \breve{J} \ . \tag{13}$$

Separating scalar and vector parts in (13) we obtain ordinary wave equations for the scalar and vector potentials

$$\frac{1}{c^2}\frac{\partial^2 \varphi}{\partial t^2} - \Delta\varphi = 4\pi\rho \ ,$$

$$\frac{1}{c^2}\frac{\partial^2 \vec{A}}{\partial t^2} - \Delta\vec{A} = \frac{4\pi}{c}\vec{j} \ .$$

On the other hand, applying in equation (11) operators $\hat{P}$ and $\hat{P}^+$ one after another to the octonic potential $\breve{\Pi}$ we can obtain first

$$\hat{P}\breve{\Pi} = \left(\frac{1}{c}\frac{\partial}{\partial t} + \vec{\nabla}\right)\left(\varphi + \vec{A}\right) = \frac{1}{c}\frac{\partial \varphi}{\partial t} + \vec{\nabla}\varphi + \frac{1}{c}\frac{\partial \vec{A}}{\partial t} + \left(\vec{\nabla}\cdot\vec{A}\right) + \left[\vec{\nabla}\times\vec{A}\right]. \tag{14}$$

We will use the standard definitions of the electric and magnetic fields in octonic form

$$\vec{E} = -\frac{1}{c}\frac{\partial \vec{A}}{\partial t} - \vec{\nabla}\varphi \ , \ \vec{H} = -i\left[\vec{\nabla}\times\vec{A}\right].$$

Taking into account Lorentz gauge condition

$$\frac{1}{c}\frac{\partial \varphi}{\partial t} + \left(\vec{\nabla}\cdot\vec{A}\right) = 0 \ ,$$

we can rewrite the result of $\hat{P}$ operation in (14) as

$$\hat{P}\breve{\Pi} = -\vec{E} + i\vec{H} \ , \tag{15}$$

where in the right part of (15) the octon of electromagnetic field $\breve{F}$ is written

$$\breve{F} = -\vec{E} + i\vec{H} \ .$$

Consequently equation (11) becomes

$$P^+\breve{F} = \breve{J} \ . \tag{16}$$

Applying the operator $\hat{P}^+$ to the octon of the electromagnetic field $\breve{F}$ we get

$$\frac{i}{c}\frac{\partial \vec{H}}{\partial t} - i\left(\vec{\nabla}\cdot\vec{H}\right) - i\left[\vec{\nabla}\times\vec{H}\right] - \frac{1}{c}\frac{\partial \vec{E}}{\partial t} + \left(\vec{\nabla}\cdot\vec{E}\right) + \left[\vec{\nabla}\times\vec{E}\right] = 4\pi\rho + \frac{4\pi}{c}\vec{j} \ . \tag{17}$$

Separating scalar, vector, pseudoscalar and pseudovector terms in equation (17), we get the system of Maxwell equations in octonic form

$$\left(\vec{\nabla}\cdot\vec{E}\right) = 4\pi\rho \qquad \text{– scalar term,}$$

$$\left[\vec{\nabla}\times\vec{E}\right] = -\frac{i}{c}\frac{\partial \vec{H}}{\partial t} \qquad \text{– pseudovector term,}$$

$$\left(\vec{\nabla}\cdot\vec{H}\right) = 0 \qquad \text{– pseudoscalar term,} \qquad (18)$$

$$\left[\vec{\nabla}\times\vec{H}\right] = \frac{4\pi i}{c}\vec{j} + \frac{i}{c}\frac{\partial \vec{E}}{\partial t} \qquad \text{– vector term.}$$

The system (18) coincides with Maxwell equations.

Applying operator $\hat{P}$ to both parts of equation (16), one can obtain the wave equation for $\vec{E}$ and $\vec{H}$ fields:

$$\left(\frac{1}{c^2}\frac{\partial^2}{\partial t^2} - \Delta\right)\left(i\vec{H} - \vec{E}\right) = \frac{4\pi}{c}\frac{\partial \rho}{\partial t} + 4\pi\vec{\nabla}\rho + \frac{4\pi}{c^2}\frac{\partial \vec{j}}{\partial t} + \frac{4\pi}{c}\left(\vec{\nabla}\cdot\vec{j}\right) + \frac{4\pi}{c}\left[\vec{\nabla}\times\vec{j}\right].$$

Separating scalar, vector, pseudoscalar and pseudovector terms we obtain the system of three equations:

$$\frac{1}{c^2}\frac{\partial^2 \vec{E}}{\partial t^2} - \Delta\vec{E} = -4\pi\vec{\nabla}\rho - \frac{4\pi}{c^2}\frac{\partial \vec{j}}{\partial t},$$

$$\frac{1}{c^2}\frac{\partial^2 \vec{H}}{\partial t^2} - \Delta\vec{H} = -\frac{4\pi}{c}i\left[\vec{\nabla}\times\vec{j}\right], \qquad (19)$$

$$\frac{\partial \rho}{\partial t} + \left(\vec{\nabla}\cdot\vec{j}\right) = 0.$$

The first two equations in (19) are the wave equations for electric and magnetic fields and the third one is the continuity equation.

## IV. RELATIONS FOR ENERGY, MOMENTUM AND LORENTZ INVARIANTS OF ELECTROMAGNETIC FIELD

The octonic algebra allows one to provide the combined calculus with different type values simultaneously. For example in this section we obtain the relations for energy, momentum and Lorentz invariants of electromagnetic field.

Multiplying both parts of expression (17) on octon $(\vec{E} + i\vec{H})$ from the left we can obtain the following octonic equation

$$\left(\vec{E} + i\vec{H}\right)\left(\frac{i}{c}\frac{\partial \vec{H}}{\partial t} - i\left(\vec{\nabla}\cdot\vec{H}\right) - i\left[\vec{\nabla}\times\vec{H}\right] - \frac{1}{c}\frac{\partial \vec{E}}{\partial t} + \left(\vec{\nabla}\cdot\vec{E}\right) + \left[\vec{\nabla}\times\vec{E}\right]\right) = \left(\vec{E} + i\vec{H}\right)\left(4\pi\rho + \frac{4\pi}{c}\vec{j}\right).$$

After multiplication we get

$$\frac{i}{c}\left(\vec{E}\cdot\frac{\partial\vec{H}}{\partial t}\right)+\frac{i}{c}\left[\vec{E}\times\frac{\partial\vec{H}}{\partial t}\right]-\frac{1}{c}\left(\vec{H}\cdot\frac{\partial\vec{H}}{\partial t}\right)-\frac{1}{c}\left[\vec{H}\times\frac{\partial\vec{H}}{\partial t}\right]-$$

$$-i\vec{E}(\vec{\nabla}\cdot\vec{H})+\vec{H}(\vec{\nabla}\cdot\vec{H})-i(\vec{E}\cdot[\vec{\nabla}\times\vec{H}])-i[\vec{E}\times[\vec{\nabla}\times\vec{H}]]+(\vec{H}\cdot[\vec{\nabla}\times\vec{H}])+[\vec{H}\times[\vec{\nabla}\times\vec{H}]]-$$

$$-\frac{1}{c}\left(\vec{E}\cdot\frac{\partial\vec{E}}{\partial t}\right)-\frac{1}{c}\left[\vec{E}\times\frac{\partial\vec{E}}{\partial t}\right]-\frac{i}{c}\left(\vec{H}\cdot\frac{\partial\vec{E}}{\partial t}\right)-\frac{i}{c}\left[\vec{H}\times\frac{\partial\vec{E}}{\partial t}\right]+ \qquad (20)$$

$$+\vec{E}(\vec{\nabla}\cdot\vec{E})+i\vec{H}(\vec{\nabla}\cdot\vec{E})+(\vec{E}\cdot[\vec{\nabla}\times\vec{E}])+[\vec{E}\times[\vec{\nabla}\times\vec{E}]]+i(\vec{H}\cdot[\vec{\nabla}\times\vec{E}])+i[\vec{H}\times[\vec{\nabla}\times\vec{E}]]=$$

$$=4\pi\rho\vec{E}+i4\pi\rho\vec{H}+\frac{4\pi}{c}(\vec{j}\cdot\vec{E})-\frac{4\pi}{c}[\vec{j}\times\vec{E}]+i\frac{4\pi}{c}(\vec{j}\cdot\vec{H})-i\frac{4\pi}{c}[\vec{j}\times\vec{H}].$$

Separating in (20) values of different type (scalar, vector, pseudoscalar and pseudovector) we obtain four relations. The scalar part of equation (20) is written

$$-\frac{1}{c}\left(\vec{E}\cdot\frac{\partial\vec{E}}{\partial t}\right)-\frac{1}{c}\left(\vec{H}\cdot\frac{\partial\vec{H}}{\partial t}\right)-i(\vec{E}\cdot[\vec{\nabla}\times\vec{H}])+i(\vec{H}\cdot[\vec{\nabla}\times\vec{E}])=\frac{4\pi}{c}(\vec{j}\cdot\vec{E}).$$

Taking into account

$$\left(\vec{H}\cdot\frac{\partial\vec{H}}{\partial t}\right)=\frac{1}{2}\frac{\partial}{\partial t}(\vec{H}\cdot\vec{H})=\frac{1}{2}\frac{\partial}{\partial t}\vec{H}^2,$$

$$\left(\vec{E}\cdot\frac{\partial\vec{E}}{\partial t}\right)=\frac{1}{2}\frac{\partial}{\partial t}(\vec{E}\cdot\vec{E})=\frac{1}{2}\frac{\partial}{\partial t}\vec{E}^2,$$

as well as

$$(\vec{\nabla}\cdot[\vec{E}\times\vec{H}])=(\vec{H}\cdot[\vec{\nabla}\times\vec{E}])-(\vec{E}\cdot[\vec{\nabla}\times\vec{H}]),$$

we obtain the following expression:

$$\frac{\partial}{\partial t}\left(\frac{\vec{E}^2+\vec{H}^2}{8\pi}\right)-\frac{c}{4\pi}i(\vec{\nabla}\cdot[\vec{E}\times\vec{H}])+(\vec{j}\cdot\vec{E})=0,$$

which is the well known Poynting theorem.

The pseudoscalar part of equation (20) is

$$\frac{i}{c}\left(\vec{E}\cdot\frac{\partial\vec{H}}{\partial t}\right)-\frac{i}{c}\left(\vec{H}\cdot\frac{\partial\vec{E}}{\partial t}\right)+(\vec{H}\cdot[\vec{\nabla}\times\vec{H}])+(\vec{E}\cdot[\vec{\nabla}\times\vec{E}])=\frac{4\pi}{c}i(\vec{j}\cdot\vec{H}). \qquad (21)$$

The expression (21) is the trivial corollary, which follows from vector and pseudovector Maxwell equations (18).

The vector part of equation (20) is

$$\frac{i}{c}\left[\vec{E}\times\frac{\partial\vec{H}}{\partial t}\right]-\frac{i}{c}\left[\vec{H}\times\frac{\partial\vec{E}}{\partial t}\right]+\vec{H}(\vec{\nabla}\cdot\vec{H})+[\vec{H}\times[\vec{\nabla}\times\vec{H}]]+\vec{E}(\vec{\nabla}\cdot\vec{E})+[\vec{E}\times[\vec{\nabla}\times\vec{E}]]=4\pi\rho\vec{E}-\frac{4\pi}{c}i[\vec{j}\times\vec{H}]. \qquad (22)$$

From (22) we obtain a well known relation between energy and momentum of the electromagnetic field

$$\vec{\nabla}\left(\frac{\vec{E}^2+\vec{H}^2}{8\pi}\right)-\frac{i}{4\pi c}\frac{\partial}{\partial t}[\vec{E}\times\vec{H}]+\rho\vec{E}-\frac{i}{c}[\vec{j}\times\vec{H}]=\frac{1}{4\pi}\{\vec{E}(\vec{\nabla}\cdot\vec{E})+(\vec{E}\cdot\vec{\nabla})\vec{E}+\vec{H}(\vec{\nabla}\cdot\vec{H})+(\vec{H}\cdot\vec{\nabla})\vec{H}\}.$$

Finally, the pseudovector part of (20) is

$$-\frac{1}{c}\left[\vec{H}\times\frac{\partial\vec{H}}{\partial t}\right]-\frac{1}{c}\left[\vec{E}\times\frac{\partial\vec{E}}{\partial t}\right]+i\vec{H}(\vec{\nabla}\cdot\vec{E})-i\vec{E}(\vec{\nabla}\cdot\vec{H})+i\left[\vec{H}\times\left[\vec{\nabla}\times\vec{E}\right]\right]-i\left[\vec{E}\times\left[\vec{\nabla}\times\vec{H}\right]\right]=4\pi i\rho\vec{H}-\frac{4\pi}{c}\left[\vec{j}\times\vec{E}\right].$$

After simple manipulations we obtain the following relation:

$$\left[\vec{H}\times\left[\vec{\nabla}\times\vec{E}\right]\right]-\left[\vec{E}\times\left[\vec{\nabla}\times\vec{H}\right]\right]+\vec{H}(\vec{\nabla}\cdot\vec{E})-\vec{E}(\vec{\nabla}\cdot\vec{H})+\frac{i}{c}\left[\vec{E}\times\frac{\partial\vec{E}}{\partial t}\right]+\frac{i}{c}\left[\vec{H}\times\frac{\partial\vec{H}}{\partial t}\right]=4\pi\rho\vec{H}+\frac{4\pi}{c}i\left[\vec{j}\times\vec{E}\right].$$

Thus in octonic algebra the simple procedure of multiplication of equation (17) on the electromagnetic field octon allows one to obtain simultaneously all the well known relations for the energy and momentum of the electromagnetic field.

On the other hand if we multiply equation (17) on the octon $(i\vec{H}-\vec{E})$ we get

$$(i\vec{H}-\vec{E})\left(\frac{i}{c}\frac{\partial\vec{H}}{\partial t}-i(\vec{\nabla}\cdot\vec{H})-i\left[\vec{\nabla}\times\vec{H}\right]-\frac{1}{c}\frac{\partial\vec{E}}{\partial t}+(\vec{\nabla}\cdot\vec{E})+\left[\vec{\nabla}\times\vec{E}\right]\right)=(i\vec{H}-\vec{E})\left(4\pi\rho+\frac{4\pi}{c}\vec{j}\right).$$

Performing multiplication we obtain

$$\begin{aligned}&-\frac{1}{c}\left(\vec{H}\cdot\frac{\partial\vec{H}}{\partial t}\right)-\frac{1}{c}\left[\vec{H}\times\frac{\partial\vec{H}}{\partial t}\right]-\frac{i}{c}\left(\vec{E}\cdot\frac{\partial\vec{H}}{\partial t}\right)-\frac{i}{c}\left[\vec{E}\times\frac{\partial\vec{H}}{\partial t}\right]+\\&+\vec{H}(\vec{\nabla}\cdot\vec{H})+i\vec{E}(\vec{\nabla}\cdot\vec{H})+\left(\vec{H}\cdot\left[\vec{\nabla}\times\vec{H}\right]\right)+\left[\vec{H}\times\left[\vec{\nabla}\times\vec{H}\right]\right]+i\left(\vec{E}\cdot\left[\vec{\nabla}\times\vec{H}\right]\right)+i\left[\vec{E}\times\left[\vec{\nabla}\times\vec{H}\right]\right]-\\&-\frac{i}{c}\left(\vec{H}\cdot\frac{\partial\vec{E}}{\partial t}\right)-\frac{i}{c}\left[\vec{H}\times\frac{\partial\vec{E}}{\partial t}\right]+\frac{1}{c}\left(\vec{E}\cdot\frac{\partial\vec{E}}{\partial t}\right)+\frac{1}{c}\left[\vec{E}\times\frac{\partial\vec{E}}{\partial t}\right]+\\&+i\vec{H}(\vec{\nabla}\cdot\vec{E})-\vec{E}(\vec{\nabla}\cdot\vec{E})+i\left(\vec{H}\cdot\left[\vec{\nabla}\times\vec{E}\right]\right)+i\left[\vec{H}\times\left[\vec{\nabla}\times\vec{E}\right]\right]-\left(\vec{E}\cdot\left[\vec{\nabla}\times\vec{E}\right]\right)-\left[\vec{E}\times\left[\vec{\nabla}\times\vec{E}\right]\right]=\\&=i4\pi\rho\vec{H}-4\pi\rho\vec{E}+i\frac{4\pi}{c}(\vec{j}\cdot\vec{H})-i\frac{4\pi}{c}\left[\vec{j}\times\vec{H}\right]-\frac{4\pi}{c}(\vec{j}\cdot\vec{E})+\frac{4\pi}{c}\left[\vec{j}\times\vec{E}\right].\end{aligned} \quad (23)$$

The scalar part of the equation (23) is written as

$$-\frac{1}{c}\left(\vec{H}\cdot\frac{\partial\vec{H}}{\partial t}\right)+\frac{1}{c}\left(\vec{E}\cdot\frac{\partial\vec{E}}{\partial t}\right)+i\left(\vec{E}\cdot\left[\vec{\nabla}\times\vec{H}\right]\right)+i\left(\vec{H}\cdot\left[\vec{\nabla}\times\vec{E}\right]\right)=-\frac{4\pi}{c}(\vec{j}\cdot\vec{E}).$$

This expression leads to the relation for the Lorentz invariant $\vec{E}^2-\vec{H}^2$, which can be represented as

$$\frac{\partial}{\partial t}\left(\frac{\vec{E}^2-\vec{H}^2}{8\pi}\right)+\frac{c}{4\pi}i\left\{\left(\vec{E}\cdot\left[\vec{\nabla}\times\vec{H}\right]\right)+\left(\vec{H}\cdot\left[\vec{\nabla}\times\vec{E}\right]\right)\right\}+(\vec{j}\cdot\vec{E})=0.$$

The pseuodoscalar part of the equation (23) is written as

$$-\frac{i}{c}\left(\vec{E}\cdot\frac{\partial\vec{H}}{\partial t}\right)-\frac{i}{c}\left(\vec{H}\cdot\frac{\partial\vec{E}}{\partial t}\right)+\left(\vec{H}\cdot\left[\vec{\nabla}\times\vec{H}\right]\right)-\left(\vec{E}\cdot\left[\vec{\nabla}\times\vec{E}\right]\right)=i\frac{4\pi}{c}(\vec{j}\cdot\vec{H}),$$

which leads to

$$\frac{1}{c}\frac{\partial}{\partial t}(\vec{E}\cdot\vec{H})+i\left(\vec{H}\cdot\left[\vec{\nabla}\times\vec{H}\right]\right)-i\left(\vec{E}\cdot\left[\vec{\nabla}\times\vec{E}\right]\right)+\frac{4\pi}{c}(\vec{j}\cdot\vec{H})=0. \quad (24)$$

The expression (24) is the relation for the second Lorentz invariant $(\vec{E}\cdot\vec{H})$.

The vector part of octonic equation (23) is written as

$$-\frac{i}{c}\left[\vec{E}\times\frac{\partial\vec{H}}{\partial t}\right]-\frac{i}{c}\left[\vec{H}\times\frac{\partial\vec{E}}{\partial t}\right]+\vec{H}(\vec{\nabla}\cdot\vec{H})-\vec{E}(\vec{\nabla}\cdot\vec{E})+\left[\vec{H}\times[\vec{\nabla}\times\vec{H}]\right]-\left[\vec{E}\times[\vec{\nabla}\times\vec{E}]\right]=-4\pi\rho\vec{E}-i\frac{4\pi}{c}\left[\vec{j}\times\vec{H}\right].$$

After transformation we obtain the following expression:

$$\vec{\nabla}\left(\frac{\vec{E}^2-\vec{H}^2}{8\pi}\right)-\frac{i}{4\pi c}\left\{\left[\vec{E}\times\frac{\partial\vec{H}}{\partial t}\right]+\left[\vec{H}\times\frac{\partial\vec{E}}{\partial t}\right]\right\}+\rho\vec{E}+\frac{i}{c}\left[\vec{j}\times\vec{H}\right]=\frac{1}{4\pi}\left\{\vec{E}(\vec{\nabla}\cdot\vec{E})+(\vec{E}\cdot\vec{\nabla})\vec{E}-\vec{H}(\vec{\nabla}\cdot\vec{H})-(\vec{H}\cdot\vec{\nabla})\vec{H}\right\}.$$

Finally, the pseudovector part of the equation (23) is given by

$$-\frac{1}{c}\left[\vec{H}\times\frac{\partial\vec{H}}{\partial t}\right]+\frac{1}{c}\left[\vec{E}\times\frac{\partial\vec{E}}{\partial t}\right]+i\vec{H}(\vec{\nabla}\cdot\vec{E})+i\vec{E}(\vec{\nabla}\cdot\vec{H})+i\left[\vec{H}\times[\vec{\nabla}\times\vec{E}]\right]+i\left[\vec{E}\times[\vec{\nabla}\times\vec{H}]\right]=4\pi i\rho\vec{H}+\frac{4\pi}{c}\left[\vec{j}\times\vec{E}\right].$$

After conversion we obtain

$$\vec{\nabla}(\vec{E}\cdot\vec{H})=\vec{H}(\vec{\nabla}\cdot\vec{E})+\vec{E}(\vec{\nabla}\cdot\vec{H})+(\vec{E}\cdot\vec{\nabla})\vec{H}+(\vec{H}\cdot\vec{\nabla})\vec{E}-4\pi\rho\vec{H}+\frac{4\pi}{c}i\left[\vec{j}\times\vec{E}\right]+\frac{i}{c}\left[\vec{H}\times\frac{\partial\vec{H}}{\partial t}\right]-\frac{i}{c}\left[\vec{E}\times\frac{\partial\vec{E}}{\partial t}\right],$$

which is the expression for the gradient of the second Lorentz invariant.

## V. OCTONIC EQUATIONS FOR THE ELECTROMAGNETIC FIELD IN A MATTER

Ordinarily electromagnetic field in a matter is described by four vectors: electric field intensity $\vec{E}$, electric induction $\vec{D}$, magnetic field intensity $\vec{H}$ and magnetic induction $\vec{B}$, which are included in the Maxwell equations in a nonsymmetrical manner:

$$\begin{aligned} div\,\vec{B} &= 0, \\ rot\,\vec{E} &= -\frac{1}{c}\frac{\partial\vec{B}}{\partial t}, \\ div\,\vec{D} &= 4\pi\rho, \\ rot\,\vec{H} &= \frac{4\pi}{c}\vec{j}+\frac{1}{c}\frac{\partial\vec{D}}{\partial t}. \end{aligned} \qquad (25)$$

Note that a simple generalization of the octonic equation (16) for electromagnetic field in a matter by introducing new octon $\breve{F}$ as the linear combination of $\vec{E}$, $\vec{D}$, $\vec{H}$ and $\vec{B}$ vectors is impossible. The action of the $\hat{P}$ operator on such octon $\breve{F}$ leads to the appearance of identical items for the $\vec{E}$, $\vec{D}$ vectors ($div\,\vec{E}$ and $div\,\vec{D}$, $rot\,\vec{E}$ and $rot\,\vec{D}$, $\partial\vec{E}/\partial t$ and $\partial\vec{D}/\partial t$) and for $\vec{H}$, $\vec{B}$ vectors as well. But equations (25) have only one item from these couples. Nevertheless the octonic equation for electromagnetic field in a matter can be formulated but in some more complicated form than equation (16).

System (25) can be divided conditionally in two pairs of independent equations for $\vec{E},\vec{B}$ vectors and for $\vec{H},\vec{D}$ vectors. Let us consider the octon $\breve{F}_{EB}=-\vec{E}+i\vec{B}$. In a vacuum this octon coincides with $\breve{F}$. The octonic equation $\hat{P}^+\breve{F}_{EB}=\breve{J}$ is similar to equation (16) and has the following form

$$\left(\frac{1}{c}\frac{\partial}{\partial t}-\vec{\nabla}\right)(i\vec{B}-\vec{E})=4\pi\rho+\frac{4\pi}{c}\vec{j}.$$

Acting by operator $\hat{P}^+$ on the octon $\breve{F}_{EB}$ we obtain

$$\frac{i}{c}\frac{\partial \vec{B}}{\partial t} - i(\vec{\nabla}\cdot\vec{B}) - i[\vec{\nabla}\times\vec{B}] - \frac{1}{c}\frac{\partial \vec{E}}{\partial t} + (\vec{\nabla}\cdot\vec{E}) + [\vec{\nabla}\times\vec{E}] = 4\pi\rho + \frac{4\pi}{c}\vec{j}.$$

In Gibbs vector form this relation looks as

$$\frac{i}{c}\frac{\partial \vec{B}}{\partial t} - i\, div\, \vec{B} + rot\, \vec{B} - \frac{1}{c}\frac{\partial \vec{E}}{\partial t} + div\, \vec{E} + i\, rot\, \vec{E} = 4\pi\rho + \frac{4\pi}{c}\vec{j}. \qquad (26)$$

Only the first, second and sixth terms in the left part of this equation are represented in equations (25). However just these terms contain the imaginary unit $i$, so they can be extracted by using complex conjugation. We will indicate complex conjugation by asterix "*". Then subtracting from (26) the conjugated equation we get octonic equation

$$\hat{P}^+\breve{F}_{EB} - \left(\hat{P}^+\breve{F}_{EB}\right)^* = 0, \qquad (27)$$

which encloses the first pair of equations (25). Indeed, separating the values of different types in (27) we obtain

$$div\, \vec{B} = 0,$$

$$rot\, \vec{E} = -\frac{1}{c}\frac{\partial \vec{B}}{\partial t}.$$

Note that $\left(\hat{P}^+\breve{F}_{EB}\right)^* \neq \hat{P}^{+*}\breve{F}_{EB}^*$, since complex conjugation changes the sign of vector product.

To obtain octonic equation corresponding to the second pair of the system (25) we will formally consider the following equation

$$\hat{P}^+\breve{F}_{DH} = \breve{J},$$

where $\breve{F}_{DH} = -\vec{D} + i\vec{H}$. If we add this equation with complex conjugated equation we get

$$\hat{P}^+\breve{F}_{DH} + \left(\hat{P}^+\breve{F}_{DH}\right)^* = 2\breve{J}, \qquad (28)$$

which encloses the second pair of Maxwell equations. Indeed, performing multiplication in (28) and separating different values we obtain

$$div\, \vec{D} = 4\pi\rho,$$

$$rot\, \vec{H} = \frac{4\pi}{c}\vec{j} + \frac{1}{c}\frac{\partial \vec{D}}{\partial t}.$$

Now it is easy to show that Maxwell equations (25) can be represented as the single generalized octonic equation for the field octon $\breve{F}_0 = i\vec{H} - i\vec{E} - \vec{B} - \vec{D}$. If we multiply the equation (27) on $i$ and add to (28) then we obtain

$$\frac{1}{2}\left\{\hat{P}^+\breve{F}_0 + \left(\hat{P}^+\breve{F}_0\right)^*\right\} = \breve{J}. \qquad (29)$$

We can rewrite equation (29) taking real part of $\hat{P}^+\breve{F}_0$:

$$Re\left\{\hat{P}^+\breve{F}_0\right\} = \breve{J}. \qquad (30)$$

So the octonic equation (30) is completely equivalent to the system of Maxwell equations in a matter.

## VI. CONCLUSION

Thus we have presented the eight-component octons (enclosing scalar, vector, pseudoscalar and pseudovector values) generating noncommutative associative algebra. On the basis of octonic algebra the generalized octonic equation for the electromagnetic field has been proposed. It was shown that this equation leads both to the wave equations for potentials and fields and to the system of Maxwell equations. The octonic equation for electromagnetic field in a matter has been also proposed.

Octonic calculus methods have been applied to the derivation of the relations for energy, momentum and Lorentz invariants of electromagnetic field. It was shown that in octonic algebra the complicated relations between values characterizing the electromagnetic field are obtained as a result of simple octonic multiplication.

The proposed octonic algebra is also convenient and natural for the generalization of the relativistic quantum mechanics equations on the basis of octonic wave functions and octonic operators that will be discussed in the next paper.

## ACKNOWLEDGEMENTS

The authors are very thankful to G.V. Mironova for kind assistance and moral support. Special thanks to Prof. C. Binns (Leicester University, UK) for useful remarks.